\documentclass[pra,aps,showpacs,twocolumn,superscriptaddress]{revtex4}
\usepackage[latin1]{inputenc}
\usepackage{eufrak}
\usepackage{amsmath}
\usepackage{xcolor}
\usepackage{graphicx}
\usepackage{subfigure}

\newcommand{\di}{\mathrm{d}}
\newcommand{\eh}{\mathrm{e}}

\newcommand{\be}{\begin{equation}}
\newcommand{\ee}{\end{equation}}
\newcommand{\bi}{\begin{itemize}}
\newcommand{\ei}{\end{itemize}}
\newcommand{\bea}{\begin{eqnarray}}
\newcommand{\eea}{\end{eqnarray}}
\newcommand{\void}[1]{}

\newcommand{\eq}[1]{Eq.~(\ref{#1})}
\newcommand{\fig}[1]{Fig.~\ref{#1}}

\renewcommand{\imath}{\mathrm{i}}

    {\catcode`\|=\active\gdef\Braket#1{\left<\mathcode`\|"8000\let|\bravert {#1}\right>}}

    \def\bravert{\egroup\,\vrule\,\bgroup}

\begin{document}
\bibliographystyle{prsty}

\title{High harmonic spectra via dominant interaction Hamiltonians}
\author{Carlos Zagoya}
\affiliation{Max Planck Institute for the Physics of Complex Systems\\
N\"othnitzer Str.\ 38, D-01187 Dresden, Germany}
\author{Christoph-Marian Goletz}
\affiliation{Institut f\"ur Theoretische Physik Technische Universit\"at 
Dresden\\D-01062 Dresden, Germany}
\author{Frank Grossmann}
\affiliation{Institut f\"ur Theoretische Physik Technische Universit\"at 
Dresden\\D-01062 Dresden, Germany}
\author{Jan-Michael Rost}
\affiliation{Max Planck Institute for the Physics of Complex Systems\\
N\"othnitzer Str.\ 38, D-01187 Dresden, Germany}
\date{\today}

\begin{abstract}
We formulate the concept of dominant interaction Hamiltonians 
to  obtain an integrable approximation to the dynamics of an electron exposed to a strong laser field and an atomic potential leading to high harmonic generation. The concept relies on local information in phase space to switch between the interactions.
This information is provided by classical integrable trajectories from which we construct a semiclassical wave function. The high harmonic spectrum obtained is in excellent agreement with the accurate quantum spectrum. The separation in the atomic potential and laser coupling interactions should facilitate the calculation of high harmonic spectra in complex systems.
\end{abstract}


\maketitle

High Harmonic Generation (HHG)  is one of the basic processes of non-linear light-matter interaction involving  an  electron under the simultaneous  influence of a strong
laser field and an atomic potential.
Initial experimental observations with atoms \cite{McPetal87,Fetal88} were soon followed by theoretical work \cite{Cor93,Letal94}, for an 
early review, see, e.~g., \cite{PKK97}.  More recently
 HHG  has been also investigated in molecules \cite{Fetal95,KLEG01} and clusters 
\cite{Hetal99,AAM00}.

The enormous impact of HHG up to recent proposals for imaging of molecular orbitals \cite{itle+04} and  the generation of attosecond pulses \cite{heki+01} is not the least due to a very simple description  with the  so-called three-step model \cite{Cor93,Letal94}.
According to this model the electron tunnels out of the combined
nuclear plus laser potential, and is then accelerated and pushed back to the nucleus by the laser field
where its energy can be converted upon recombination
into a high energy photon. This is possible up to a maximal  cutoff energy
of $E_{\mathrm{max}}= $3.17$U_p+I_p$, where $U_p ={\cal E}^{2}/(4\omega^{2})$ is the ponderomotive energy in terms of the 
maximal laser field strength ${\cal E}$ and the photon frequency $\omega$, while  
$I_p$ the ionization potential. $E_{\mathrm{max}}$ was derived by considering classically the excursion of the electron 
in the laser field only \cite{Cor93}.  On the other hand in purely classical HHG calculations the cutoff energy does not 
play a role since HHG  is  a quantum mechanical interference effect, essentially between electronic quantum amplitude 
in the ground state and in the continuum driven by the laser, as the three step model reveals qualitatively. With a full semiclassical calculation based on trajectories in time, excellent agreement with the quantum result is achieved \cite{vdSR992}. However,  this also implies that a quantitative description of HHG can only be achieved by solving numerically the non-separable dynamics of an electron in the  Coulomb field of a nucleus with potential $V_{a}$  and in the external laser field  $(V_{{\cal E}})$.

In view of the universal relevance of HHG as mentioned above it would be very desirable to have an analytical approach which is also quantitative. To this end
we introduce the idea of {\it dominant interaction Hamiltonians}  (DIH) to disentangle the non-separable dynamics by splitting it into  spatial regions where either one of the two potentials dominates which is then taken as the only interaction in that region.
For regions where the laser dominates (far away from the nucleus) the dynamics is simply that of a free electron in the laser field, the so called Volkov-dynamics, governed by the Hamiltonian $H_{{\cal E}} = p^{2}/2 + V_{{\cal E}}$. In the opposite case we have the electron only under the influence of the Coulomb potential $V_{a}$ which is trivially integrable for one electron with $H_{a}= p^{2}/2 + V_{a}$.
  
\begin{figure}[t]
\noindent \begin{centering}
\includegraphics[width=.85\columnwidth]{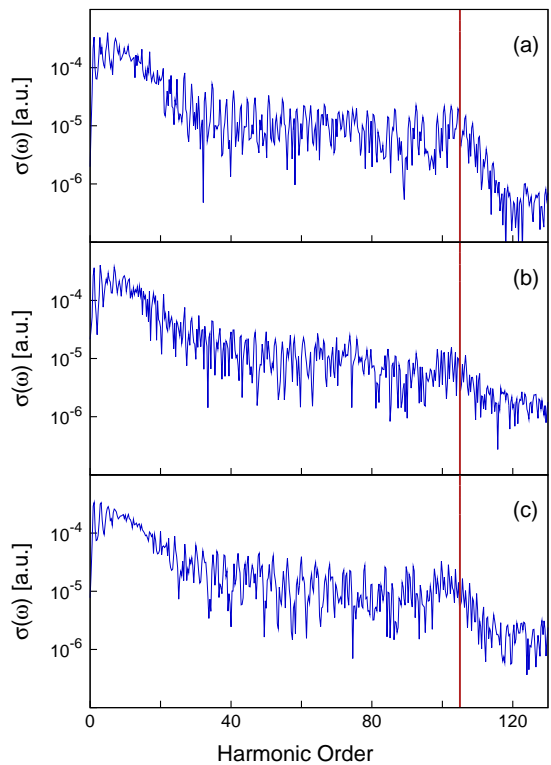}
\par\end{centering}
\noindent \begin{raggedright}
\caption{\label{fig:multispectrum} (color online) HH spectrum calculated from
Fourier transformation of the dipole acceleration, (a) full quantum result, (b) 
semiclassical result for the full potential with $10^6$ trajectories, (c)  DIH result with dominant interactions   
with $10^4$ trajectories. The cutoff for scattering from an ion under a laser field, $E_\mathrm{cutoff}=2 U_{p}+I_{p}$,  
determines the highest harmonic order, $N_{\mathrm{max}}=\lfloor E_\mathrm{cutoff}/\omega\rfloor=105$. It is shown with a 
vertical (red) line.}
\par\end{raggedright}
\end{figure}

\begin{figure}[t]
\noindent \begin{centering}
\includegraphics[width=.85\columnwidth]{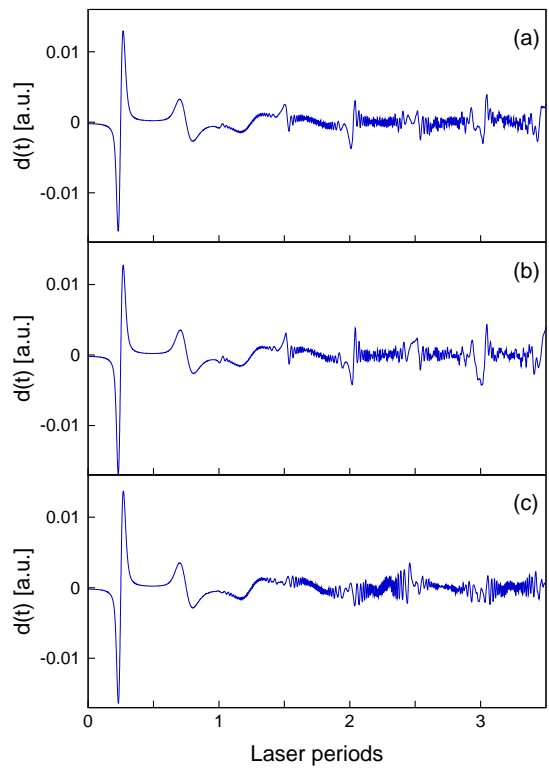}
\par\end{centering}
\noindent \begin{raggedright}
\caption{\label{fig:multiaccel} Dipole acceleration $d(t)$, from which the HH spectrum 
shown in \fig{fig:multispectrum} has been obtained.}
\par\end{raggedright} 
\end{figure}

To demonstrate how the concept of DIH works, we will use the simplest realization of HHG,
a one dimensional electron dynamics in a soft core potential $V_{a}$ along the linearly polarized laser field,
defined by the Hamiltonian (atomic units (a.u.) are used unless stated otherwise)
 \be\label{ham}
 H = \frac{p^{2}}2 + V_{a}+V_{{\cal E}},
 \ee
where $V_{a}(x)= -(x^{2}+a)^{-1/2}$ with $a=2$~a.u.\ such that the ground 
state energy  agrees with hydrogen ($-E_{b}=1/2=I_p$). The laser interaction 
is defined as $V_{{\cal E}}= x {\cal E}\cos\omega t$,
where for convenience we use here a 3.5-cycle laser pulse with 
${\cal E} = $0.1~a.u.\ and $\omega = 0.0378$~a.u. \cite{vdSR992}. Our observable of interest, the HH spectrum 
$\sigma(\omega)$ can formulated as the Fourier transform  $\sigma(\omega) = \int \di t\,d(t) \exp[\imath \omega t]$ of the 
dipole acceleration
 \be\label{dipole}
 d(t) = -\langle \Psi(t)|\partial V_{a}/\partial x|\Psi(t)\rangle\,.
 \ee
 The exact  $\Psi(t)$ is the solution of the time-dependent Schr\"odinger equation with Hamiltonian
\eq{ham} and leads to the familiar HH spectrum shown in \fig{fig:multispectrum} (a)
which was extracted from the acceleration in \fig{fig:multiaccel} (a). It has been obtained under scattering conditions, i.e., with a Gaussian electron wavepacket  $\Psi_{i}(x,0)=
\langle x|g(\mathbf w_{i})\rangle$, where
\begin{multline}
\label{eq:wp}
\langle x|g(\mathbf w_{i})\rangle\equiv
\left( \frac{\gamma}{\pi} \right)^{\frac 14} 
\exp\bigg\{-\frac{\gamma}{2}(x-q_{i})^2 
+ {\imath}p_{i}(x-q_{i})\bigg\}\,.
\end{multline}
We have introduced the short notation $\mathbf w = (p,q)$ for a point in phase space.
The wavepacket  $\Psi_{i}$  with width parameter $\gamma=0.05$~a.u.. is initially located at rest ($p_{i}=0$) at a 
distance of $q_{i}= {\cal E}/\omega^{2}$ (70 a.u.) from the proton  which corresponds to half the quiver amplitude.


\begin{figure}[t]
\noindent \begin{centering}
\includegraphics[angle=-90, scale=0.35]{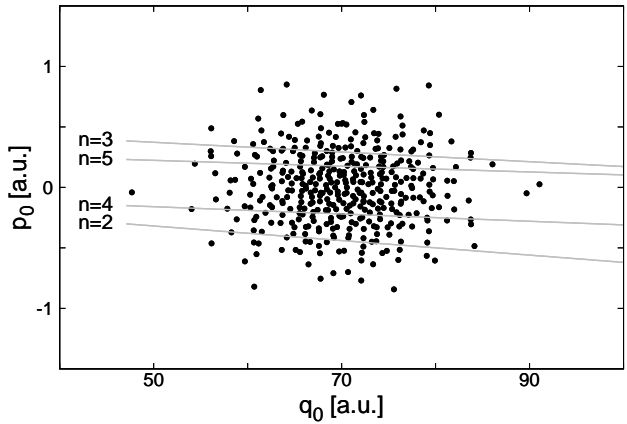}
\par\end{centering}
\noindent \begin{raggedright}
\caption{\label{fig:initial2} Distribution of initial conditions  (dots) 
and analytical conditions for switching (lines, see Eq.~(\ref{eq:stripes})).}
\par\end{raggedright}
\end{figure}

Next,  we construct the semiclassical  HH spectrum. While the full quantum spectrum only serves as a reference for accuracy of our approximation, the semiclassical propagation is part of the DIH approach to be developed since the latter requires local information in phase space as we will see.
Such information is contained in the classical trajectories underlying the semiclassical propagator
developed by Herman and 
Kluk \cite{HK84}, (see also 
 \cite{He81}  and \cite{Kay941}),
\begin{multline}
K(x,x',t) =\int 
\frac{\di^2w_{0}}{2\pi}\; \langle x|g(\mathbf w_{t})\rangle 
\sqrt{R}\;\eh^{\imath S}\; 
\langle g(\mathbf w_{0})|x'\rangle\,,
\label{eq:hk}
\end{multline}
with Gaussians $\langle x|g(\mathbf w)\rangle$ which have for convenience the same width as the initial state \eq{eq:wp}.
The interpretation of \eq{eq:hk} is straight forward: The quantum transition amplitude
from point $x'$ at time $t=0$ to $x$ at time $t$ is constructed
through classical trajectories which start at $\mathbf w_{0}$ at time $t=0$ and reach
under the dynamics of the Hamiltonian the phase space point $\mathbf w_{t}$ at time $t$.
The preexponential weight factor of such a trajectory in phase space is given by 
\begin{multline}
R=\frac{1}{2}\det\bigg(m_{p_{t}p_{0}}+m_{q_{t}q_{0}} 
-\imath \gamma m_{q_{t}p_{0}}+\frac{\imath}{\gamma}m_{p_{t}q_{0}}\bigg)\,,
\label{prefac}
\end{multline}
which is composed out of the four blocks  $m_{ab}\equiv \partial a/\partial b$
of the monodromy matrix \cite{HK84}. Note that for the present one-dimensional case,
the $m_{ab}$ are scalars and no determinant has to be taken. The semiclassical amplitude is then given by 
$\sqrt R e^{\imath S}$, where $S(t)$ is the action along the trajectory.  The  integration is performed over all 
phase space points $\mathbf w_{0}$ which serve as initial conditions of classical trajectories 
$\mathbf w_{t} \equiv (p_t=p(p_0,q_0,t),q_t=q(p_0,q_0,t))$. Convergence
is achieved  with a finite number of trajectories through the Gaussian envelope, as illustrated in \fig{fig:initial2}. 
The HH spectrum obtained with the semiclassical wavefunction
$\psi(x,t) = \int \di x K(x,x',t)\psi_{i}(x')$ is in excellent agreement with the quantum spectrum, 
see \fig{fig:multispectrum}. While providing a lot of insight into the dynamics which creates HH  the semiclassical 
approach is at least as numerically involved as solving the Schr\"odinger equation directly, since the trajectories 
$\mathbf w_{t}$ cannot be obtained analytically, and moreover, the full classical dynamics of this problem is chaotic \cite{vdSR992}, although the HH spectrum is very regular. 

\begin{figure}[t]
\noindent \begin{centering}
\includegraphics[width=.75\columnwidth]{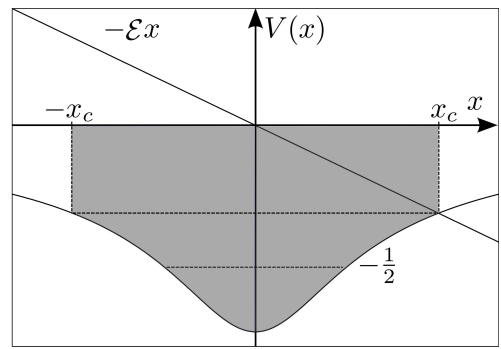}
\par\end{centering}
\noindent \begin{raggedright}
\caption{\label{fig:sw-scheme} Soft Coulomb $V_{a}$ and laser potential $-x {\cal E}$. Shaded areas represent the switching regions.}
\par\end{raggedright}
\end{figure}

This underscores the motivation for the DIH concept, where for each dominant interaction, the trajectories are ideally 
known analytically, or at least can be obtained with little numerical effort.  The key point of the DIH approach is to 
define an appropriate phase space boundary between  the electron dynamics governed by the atomic potential $V_{a}$ and 
the laser potential $V_{{\cal E}}$.  For the present one dimensional case, the boundary reduces to isolated points 
$\mathbf w_{c}$ in phase space. The physical process we have to describe is the trapping of the  freely oscillating 
electron in the laser field, due to the atomic potential. This will be most likely if the electron is slow close to 
the nucleus, i.e., we set $p_{c}=0$. The trapping region is defined in a natural way as the range where the soft-core 
potential is always stronger than the laser potential. For $V_a$ and $V_{\cal E}$ from \eq{ham}, this region 
results to be the interval $[-x_c,\,x_c]$, with $x_c=3.0083$ (see \fig{fig:sw-scheme}). 

We are now in a position to generate the HH spectrum with the DIH approach. Although this can be 
done completely analytically \cite{zago+12a},  we prefer to use here the same propagation scheme of classical trajectories as used for the semiclassical HH spectrum presented above. This allows for strict comparison  of the full semiclassical (and quantum)  spectrum and the one to be calculated with DIH. The DIH result agrees remarkably well with the exact quantum spectrum as can  be seen in \fig{fig:multispectrum} and \fig{fig:multiaccel} .

\begin{figure}[t]
\noindent \begin{centering}
\includegraphics[angle=-90, scale=0.35]{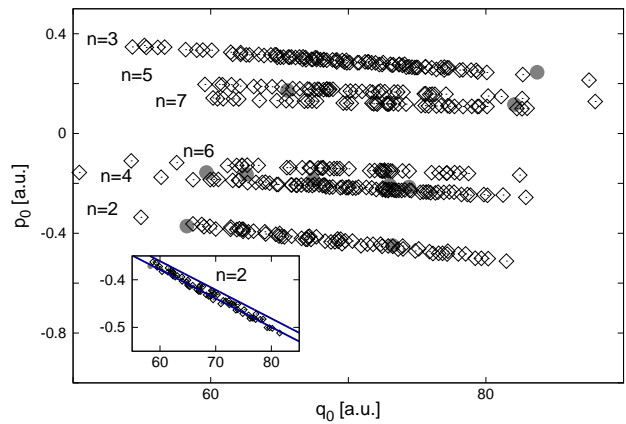}
\par\end{centering}
\noindent \begin{raggedright}
\caption{\label{fig:initial} Initial conditions for trajectories: switched 
trajectories (diamonds), trapped or stranded trajectories in the full 
potential case (filled circles) and analytical conditions (lines in the inset, see Eq.~(\ref{eq:stripes})).}
\par\end{raggedright}
\end{figure} 
 
In order to understand why the DIH approach works so well, it is instructive to analyze the initial conditions of the trajectories which switch and get trapped. A little thought reveals that they form bands in the initial phase space (see \fig{fig:initial2}). To see that we recall that the trajectory for an electron in a laser field with initial conditions $q(0)=q_0$ and $p(0)=p_0$ reads
\begin{eqnarray}\label{eq:motion_equations}
p(t) &=& p_0 - \frac{{\cal E}}{\omega}\sin\left(\omega t\right), 
\\ 
q(t) &=& q_0 + p_0t + \frac{{\cal E}}{\omega^2}\left[\cos\left(\omega t\right) 
- 1\right] \, .
\end{eqnarray}
%
From the condition  $p(t_c)=0$ follows that $t_c\approx n\pi/\omega$ with 
$n=1,\,2,\dots$. Then, the conditions for a switch from $V_{{\cal E}}$ 
to $V_{a}$ are 
%
\begin{align}
q_c= & q_0+\frac{n\pi}{\omega}p_0+\frac{{\cal E}}{\omega^2}\left[\cos\left(n\pi\right)-1\right]  \nonumber \\
= & q_0+\frac{n\pi}{\omega}p_0 - \frac{1-(-1)^{n}}{2}\frac{2{\cal E}}{\omega^2}\,.
\label{eq:switch-cond1}
\end{align}
This implies that the initial phase space points are given by lines $p^{(n)}_{0}(q_{0})$ with a width
$\Delta p^{(n)}_{0}=\omega/(n\pi)$ as illustrated in Figs. \ref{fig:initial2} and \ref{fig:initial}. 
The explicit expression follows from rearranging Eq.\ \eqref{eq:switch-cond1}, 
%
\begin{equation}\label{eq:stripes}
  p^{(n)}_{0}(q_0)=-\frac{\omega}{n\pi}\left(q_0-\frac{1-(-1)^{n}}{2}\frac{2{\cal E}}{\omega^2}\right)\,.
\end{equation}

Interestingly, for the full interaction, the initial conditions for those  trajectories which get trapped lie on the same 
phase space stripes (\fig{fig:initial}). We may conclude that the DIH switching condition describes the dynamics relevant 
for HHG quite well. The small differences in the initial conditions can be attributed to the  (small) attraction by the 
nucleus which the electrons from the full classical trajectories feel on the way inward.

This observation also explains the long standing puzzle why the Coulomb long range nature of the potential plays only a 
minor role: In fact, HH spectra in qualitative agreement with experiments have been also calculated with zero-range potentials 
$V_{a}\propto \delta(x)$ \cite{BLM94}. In the DIH approach, the switching condition would change to  $x_{c}=0$ but the 
structure of the initial manifold leading to switching remains the same. 

To summarize, we have introduced the concept of dominant interaction Hamiltonians (DIH) to
simplify the theoretical description of high harmonic generation by splitting the problem into two integrable ones: the 
electron under the influence of the laser field and the electron under the influence of the atomic potential. We construct the 
HH spectrum semiclassically by using classical trajectories: They feel the force of the laser {\it or} of the atomic potential 
and the force is switched at the phase space boundary defining the dominance of each of the two interactions.
The dynamics is integrable under either of the two interactions reducing greatly its complexity without loss of accuracy of 
the spectrum. The simplification manifests itself in the fact that up to one million trajectories are necessary in the present 
example of HHG to converge the spectrum fully semiclassically while a factor of 100 less is sufficient to converge the DIH spectrum.

Moreover, the DIH provides a natural dynamical extension of the simple man's approach: In the latter the wave functions of the 
electron in the laser field  and the electron in the ground state of the atomic potential only are simply coherently added to 
produce qualitatively the HHG spectrum \cite{vdSR00}. Here, with the help of DIH we have provided a framework how to 
dynamically populate one of these states (the ground state) while starting initially with the other one.

While the HH spectrum presented here can be obtained fully analytically  \cite{zago+12a}, the main thrust of the DIH lies 
in the perspective to describe HHG in more complex systems with the simplification of separating the interactions.
Presently, one has to resort to semiclassical techniques since the switching condition is local in phase space.  
We will explore the possibility of a corresponding quantum condition in future work.

%
FG and CMG would like to thank the Deutsche Forschungsgemeinschaft for 
financial support through grant GR 1210/4-2.
%

	
\end{document}